\documentclass{naturep}
\usepackage[utf8]{inputenc}
\usepackage{lineno}
\usepackage{setspace}
\usepackage{color}
\usepackage{graphicx}
\usepackage{booktabs} 
\usepackage{adjustbox}
\usepackage[margin=0.75in]{geometry}
\usepackage{verbatim}

\newcommand{\andrew}[1]{\textcolor{black}{#1}}

\title{\textbf{Artificial Intelligence for Digital and Computational Pathology}}
\author{Andrew H. Song$^{1,2,3,4,8}$, Guillaume Jaume$^{1,2,3,4,8}$, Drew F.K. Williamson$^{1,2,3,7}$, Ming Y. Lu$^{1,2,3,4,5}$, Anurag Vaidya$^{1,2,3,4,6}$, Tiffany R. Miller$^{1,7}$, and Faisal Mahmood$^{1,2,3,4,\ast}$}
\date{}

\makeatletter
\let\saved@includegraphics\includegraphics
\AtBeginDocument{\let\includegraphics\saved@includegraphics}
\renewenvironment*{figure}{\@float{figure}}{\end@float}
\makeatother

\begin{document}
\maketitle

\begin{spacing}{1.32}

\begin{affiliations}
 \item Department of Pathology, Brigham and Women's Hospital, Harvard Medical School, Boston, MA, USA
 \item Department of Pathology, Massachusetts General Hospital, Harvard Medical School, Boston, MA, USA
 \item Cancer Program, Broad Institute of Harvard and MIT, Cambridge, MA, USA
 \item Data Science Program, Dana-Farber Cancer Institute, Boston, MA, USA
 \item Department of Electrical Engineering and Computer Science, Massachusetts Institute of Technology, Cambridge, MA, USA
 \item Harvard-MIT Division of Health Sciences and Technology, Massachusetts Institute of Technology, Cambridge, MA, USA
 \item Harvard Medical School, Boston, MA, USA
 \item These authors contributed equally.
 \end{affiliations}

\noindent$\ast$\textbf{Correspondence:}\\
Faisal Mahmood\\
60 Fenwood Road, Hale Building for Transformative Medicine\\
Brigham and Women’s Hospital, Harvard Medical School\\
Boston, MA 02445\\
faisalmahmood@bwh.harvard.edu

\clearpage
\section*{Abstract}

\begin{abstract}
    {\normalfont Advances in digitizing tissue slides and the fast-paced progress in artificial intelligence, including deep learning, have boosted the field of computational pathology. This field holds tremendous potential to automate clinical diagnosis, predict patient prognosis and response to therapy, and discover new morphological biomarkers from tissue images. Some of these artificial intelligence-based systems are now getting approved to assist clinical diagnosis; however, technical barriers remain for their widespread clinical adoption and integration as a research tool. This Review consolidates recent methodological advances in computational pathology for predicting clinical end points in whole-slide images and highlights how these developments enable the automation of clinical practice and the discovery of new biomarkers. We then provide future perspectives as the field expands into a broader range of clinical and research tasks with increasingly diverse modalities of clinical data.}
\end{abstract}

\section*{Key points}
\begin{itemize}
    \item Supported by advances in artificial intelligence, curation of multi-institutional cohorts and the development of high-performance computing, computational pathology is now reaching clinical-grade performance for certain tasks.
    
    \item Artificial intelligence-based methods in computational pathology can be distinguished into methods for predicting clinical end points from tissue specimens and assistive tools for clinical or research tasks.
    
    \item Multiple instance learning is a rapidly growing paradigm for predicting clinical endpoints, such as disease diagnosis and molecular alterations, from whole-slide images.
    
    \item Computational pathology can be used for automating tasks that pathologists already perform in daily practice and for discovering morphological biomarkers for clinical outcomes of interest.
    
    \item Initiatives for collecting larger, well-curated, and multimodal datasets and advances in AI frameworks are required for computational pathology applications to get closer to clinical adoption. 
    
\end{itemize}
 

\clearpage
\section{Introduction}

\begin{figure*}
    \centering
    \includegraphics[width=0.9\textwidth]{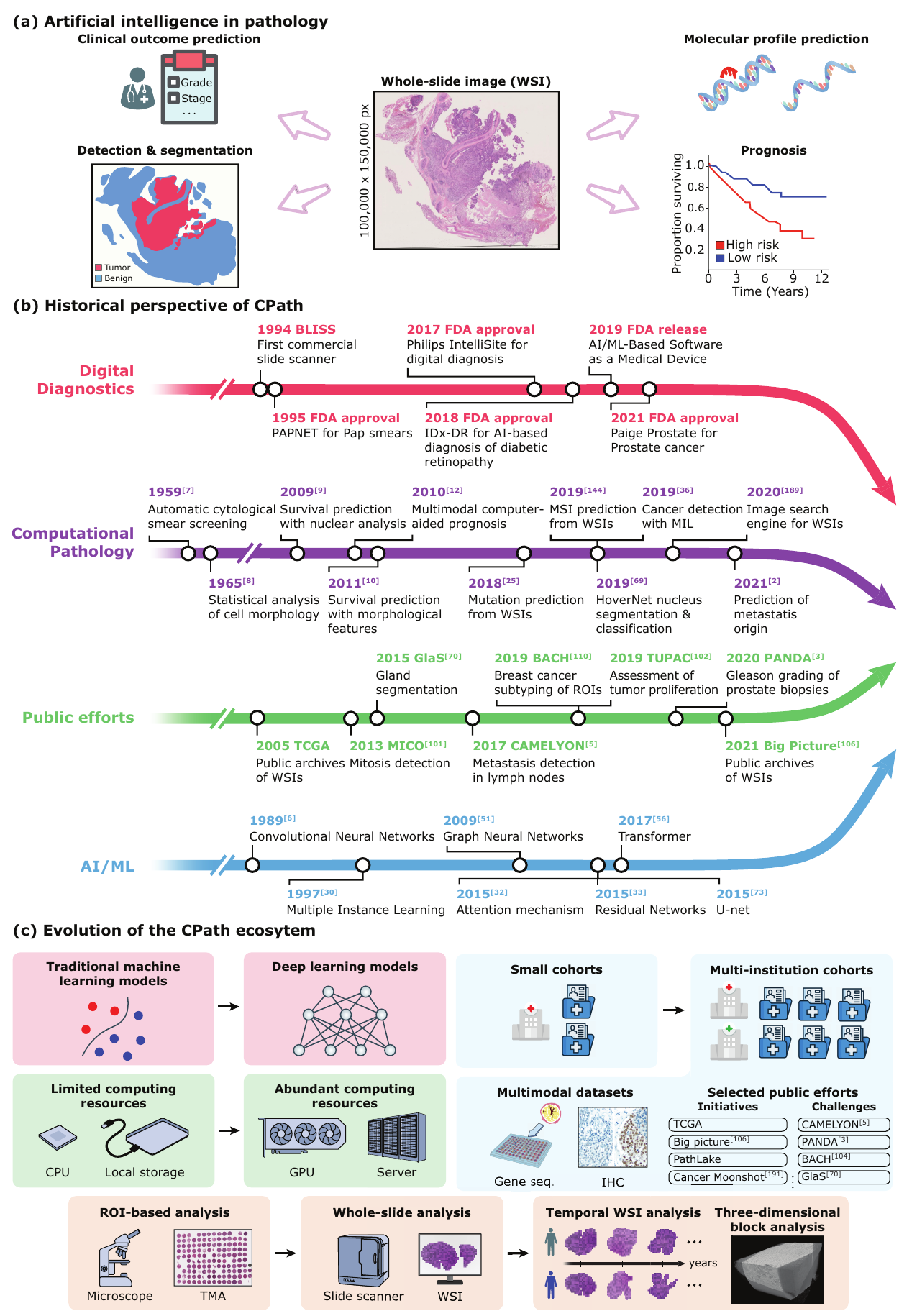}
    \caption{\textbf{Caption next page.}}
    \label{fig:timeline}
\end{figure*}

Advances in scanning systems, imaging technologies and storage devices are generating an ever-increasing volume of whole-slide images (WSIs) acquired in clinical facilities, which can be computationally analysed using artificial intelligence (AI) and deep learning technologies. The digitization and automation of clinical pathology, also referred to as computational pathology (CPath), can provide patients and clinicians the means for more objective diagnoses and prognoses, allows the discovery of novel biomarkers and can help to predict response to therapy\cite{abels2019computational}. For example, developments in AI-assisted diagnosis and automatic classification of haematoxylin and eosin (H\&E)-stained WSIs can help to determine the origin of a cancer of unknown primary\cite{lu2021ai}, grade prostate cancer on par with experienced pathologists\cite{bulten2022artificial}, predict the prognosis of patients with colorectal cancer better than conventional cancer stages\cite{skrede2020deep} and detect lymph node metastases in breast cancer\cite{bejnordi2017diagnostic} (Fig. \ref{fig:timeline} (a)).

Institutions are now gathering massive repositories of digitized slides, either with the integration of slide scanning into the routine workflow or by digitizing slide archives. The collection of large cohorts on different disease models has also been supported by public efforts, such as The Cancer Genome Atlas (TCGA) Program by the National Cancer Institute. The drastic reduction in computer storage costs and the availability of more computationally capable processors — especially graphics processing units (GPUs) — now allow laboratories to run large-scale studies based on thousands of samples. Moreover, building on the success of computer vision\cite{lecun2015deep}, AI and deep learning integration in CPath has advanced to such a degree that deep learning can now be considered the central algorithmic component of most CPath systems. These developments are major improvements for a field that began with simple statistical analyses of nuclear morphology in the 1960s\cite{bostrom1959instrumentation, prewitt1966analysis} and that today aims to transform the clinical practice of pathology (Fig. \ref{fig:timeline} (b,c)). Pioneering works based on machine learning have shown that hand-crafted, human-interpretable features (HIFs) extracted from regions of interest can be used to derive valuable diagnostic and prognostic information\cite{fuchs2008computational,beck2011systematic,madabhushi2016image}. These principles are now being scaled with deep learning, which can automatically identify and extract relevant morphological features from high-dimensional input data.

Although mainly regarded as a tool to automate specific tasks to reduce inter-observer variability or alleviate the burden on the pathologist, CPath can also help to discover new biomarkers\cite{tarantino2021evolving}. For example, through tissue analysis, AI can help to decipher different biological phenomena\cite{lee2022derivation} and identify new morphological features relevant for diagnosis and prognosis\cite{saltz2018spatial, marusyk2020intratumor}. Beyond clinical settings, CPath can be used for therapy and drug development\cite{vamathevan2019applications} by automating the identification of morphological changes in tissue specimens upon drug exposure in preclinical and clinical trials.

Despite these advances, CPath has room to grow in pathology research and precision medicine. In particular, advances in computer vision research are continuously providing new methods to improve CPath algorithms, for example, in representation learning with vision transformers\cite{dosovitskiy2020image} and self-supervised learning (SSL)\cite{krishnan2022self}. Furthermore, the shift to precision medicine requires an ever-increasing number of assays\cite{jiang2022big}, which has boosted the amount of data collected per patient that can be integrated into the CPath workflow. For example, diagnosis is no longer solely based on the histological analysis of the tissue but is also complemented by molecular and immunohistochemical assays. The emergence of multiplex imaging\cite{andreou2022multiplexed}, spatially resolved genomic assays\cite{marx2021method} and 3D pathology\cite{liu2021harnessing}, among other methodologies, will only accelerate this trend, providing new opportunities for multimodal integration.

Understanding how and to which AI frameworks the computational pathology community will converge is crucial to anticipating future challenges. In this spirit, this Review aims to identify and consolidate the major technical developments for WSI modeling. Moreover, we outline promising research areas focusing on building robust and generalizable representations of WSIs from large-scale, diverse, multimodal, and privacy-preserving datasets.

\setcounter{figure}{0}    
\begin{figure}
    \centering
    \caption{\textbf{Applications, timeline of selected milestones and trends in computational pathology.} \textbf{a}, Overview of computational pathology (CPath) applications. \textbf{b}, Digital diagnostics and artificial intelligence (AI) have made considerable progress over the past decades, laying the foundations for CPath to make a clinical impact. The timelines include selected milestones that have substantially impacted CPath. \textbf{c}, CPath shifted from traditional machine learning (ML) models based on small cohorts of regions of interest (ROIs) to deep learning models trained on large, sometimes multimodal, multi-institutional cohorts of whole-slide images (WSIs). Higher-dimensional pathology data, such as WSIs collected longitudinally for each patient and 3D tissue images, are also expected to gain traction. The digitization of the pathology workflow, abundant computational resources, public datasets and advances in AI and computer vision have supported this transition. CPU, central processing unit; gene-seq, gene sequencing; GPU, graphics processing unit; IHC, immunohistochemistry; MIL, multiple instance learning; MSI, microsatellite instability; TCGA, The Cancer Genome Atlas; TMA, tissue microarray.}
    \label{fig:timeline2}
\end{figure}

\section{Deep learning in CPath}

The methodological contributions of deep learning in CPath can be distinguished into approaches for predicting clinical end points, such as cancer subtype, patient survival or genetic mutations from WSIs, and AI-based assistive tools to guide and provide support to pathologists and researchers, such as methods for segmenting images or virtual staining (Fig. \ref{fig:attention}). 

\subsubsection*{2.1 Tissue pre-processing}

The digitization of histology slides consists of building a pyramidal structure of the tissue by representing it as images at multiple magnifications (or resolutions), typically ranging from $\times$40 ($\sim0.25\, \mu$m/pixel resolution) to $\times$5 magnification ($\sim2\, \mu$m/pixel resolution). Before any AI algorithm is applied, the digitized WSIs undergo tissue segmentation to remove background regions either by classical image processing (such as image thresholding) or by a deep learning-based approach (such as segmentation networks\cite{bandi2018detection}). Due to the substantial dimensions of WSIs (a WSI can be up to 100,000$\times$ larger than an ImageNet sample of 256$\times$256 pixels\cite{deng2009imagenet}), direct processing is computationally demanding. Consequently, it is common practice to partition WSIs into small patches. With patching, CPath frameworks can adopt a divide-and-conquer strategy, in which each patch is individually processed with a neural network, the results of which can be further aggregated to yield a slide-level and/or patient-level outcome. At high magnification (for example, $\times$40 or $\times$20), each image patch reveals granular information, such as nuclear morphology. However, only a small context is visible, which may limit the ability of a model to capture large contextual patterns. Conversely, processing at lower magnifications (for example, $\times$10 or $\times$5) provides more contextual information per patch, such as tissue architecture, although at the cost of reduced resolution. Therefore, context and resolution need to be balanced based on the specific application; for example, certain cancer subtyping tasks, such as lung carcinoma subtyping, can be performed at $\times$5 magnification with clinical-grade accuracy, whereas genetic mutation prediction from a WSI usually requires $\times$20 magnification or higher\cite{coudray2018classification}. 

\subsubsection*{2.2 Multiple instance learning on WSI}

One of the main objectives of CPath is to predict disease-related clinical endpoints from WSIs -- a task referred to as WSI classification.

One way to address the computational bottleneck associated with the large WSI size and, in turn, the large number of patches, is to reformulate slide classification as a patch-level supervised learning classification task. This approach involves processing each patch using a feature extractor, such as a convolutional neural network (CNN), to obtain a patch embedding. The resulting embedding is then passed to a predictor for predicting the corresponding patch label. The labels can be provided by pathologists by manually annotating regions of interest or by assigning the same label to all patches in the slide. After patch classification, patch-level scores are combined using an aggregator to make a WSI-level prediction. The aggregator can be non-parameterized, such as taking the average, maximum or majority voting, or parameterized with an additional neural network. However, the patch-level supervision approach has several limitations. First, obtaining manual annotations is time-consuming and cannot be easily scaled to thousands of WSIs. Moreover, the meaning of a patch label becomes ambiguous for applications such as prognosis or therapy response prediction, for which pathologists have minimal a priori knowledge. Intratumoural heterogeneity\cite{marusyk12intra,vitale2021intratumoral} further complicates assigning annotations even within the same tumour region. Furthermore, assigning the same label to all patches in the WSI works if the region of interest occupies most of the WSI\cite{coudray2018classification,kather2020pan}. However, when only a fraction of the image is discriminative (for example, lymph-node metastases), the patch-level labels become very noisy\cite{ilse18attention} (Fig.~\ref{fig:attention} (a)).

Alternatively, WSI classification can be defined with multiple instance learning (MIL)\cite{dietterich1997solving,dundar2010multiple}, in which a single supervisory label is provided to the set of patches constituting the WSI and only a subset of the patches is assumed to correspond to that label (Fig.~\ref{fig:attention} (b)). This setting is also referred to as weakly supervised learning because the number of patches is substantially larger than the number of supervisory labels. MIL learns to map the set of patches to the labels in three steps: first, a feature extractor extracts a low-dimensional embedding of each patch (for example, a 1,024-dimensional embedding); second, an aggregator pools the patch embeddings to form a WSI representation; and third, the said representation is mapped to the WSI label using a predictor. MIL differs from patch-level learning in that the WSI-level label is no longer assigned to patches but to the set of patches constituting the WSI. The aggregator can be a non-parameterized function, such as the average or maximum of the patch embeddings, or a parametrized function, the most popular being the attention mechanism. In attention-based MIL\cite{ilse18attention,lu2020deep,lu2021ai}, each patch is assigned an attention score based on the importance of its embedding towards rendering the prediction, which is further used to derive an attention-weighted sum of the patch embeddings. One advantage of attention-based MIL is that these attention scores can directly provide interpretable heatmaps for qualitative morphological analysis.

However, because of the large number of patches per WSI, the entire set of WSI patches and the entire network cannot be stored in GPU memory simultaneously. Therefore, MIL cannot readily learn the feature extractor and the predictor in a joint manner. One solution to this problem is to pre-train the feature extractor on an auxiliary task to pre-extract patch embeddings. By this approach, the aggregator and the predictor operate on patch embeddings that are already compressed (for example, 1,024-dimensional embeddings compress 256 $\times$ 256 patches by a factor of around 200). In practice, the feature extractor can be pre-trained on natural image datasets such as ImageNet\cite{deng2009imagenet} (for example using a ResNet\cite{he2016deep}), on histopathology images with an auxiliary task\cite{shaban20,tellez2018whole,campanella2019clinical} or SSL\cite{wang2022transformer,ciga2022self,chen2022scaling}. Differently, engineering steps optimizing the available memory by either accessing the host memory\cite{chen2021annotation} or using gradient checkpointing\cite{pinckaers2020streaming,huang2022deep} can be used for joint feature extractor and predictor training. Nonetheless, these strategies are complex and computationally burdensome, so they have not been widely adopted so far. Another approach is to randomly sample a subset of the WSI patches during training\cite{wulczyn2020deep,wulczyn2021interpretable}, under the assumption that relevant information gets sampled each time.

Overcoming computational constraints and adapting them to an increasingly wide range of CPath tasks will continue to influence model design. Although a recent benchmark showed that patch-level training and MIL can perform similarly for a range of tasks\cite{laleh2022benchmarking}, such as tumour subtyping, we argue that MIL methods will gain traction. First, the field is evolving towards more complex tasks with weak training signals (such as survival prediction), for which patch-level supervision is not suitable. Second, new hardware advances will eventually allow for the joint training of the feature extractor, aggregator and predictor with the entire set of WSI patches in a simple, off-the-shelf manner, enabling MIL to incorporate more contextual information.

\subsubsection*{2.2 Emergence of context-aware approaches}

\begin{figure}
    \centering
    \includegraphics[width=\textwidth]{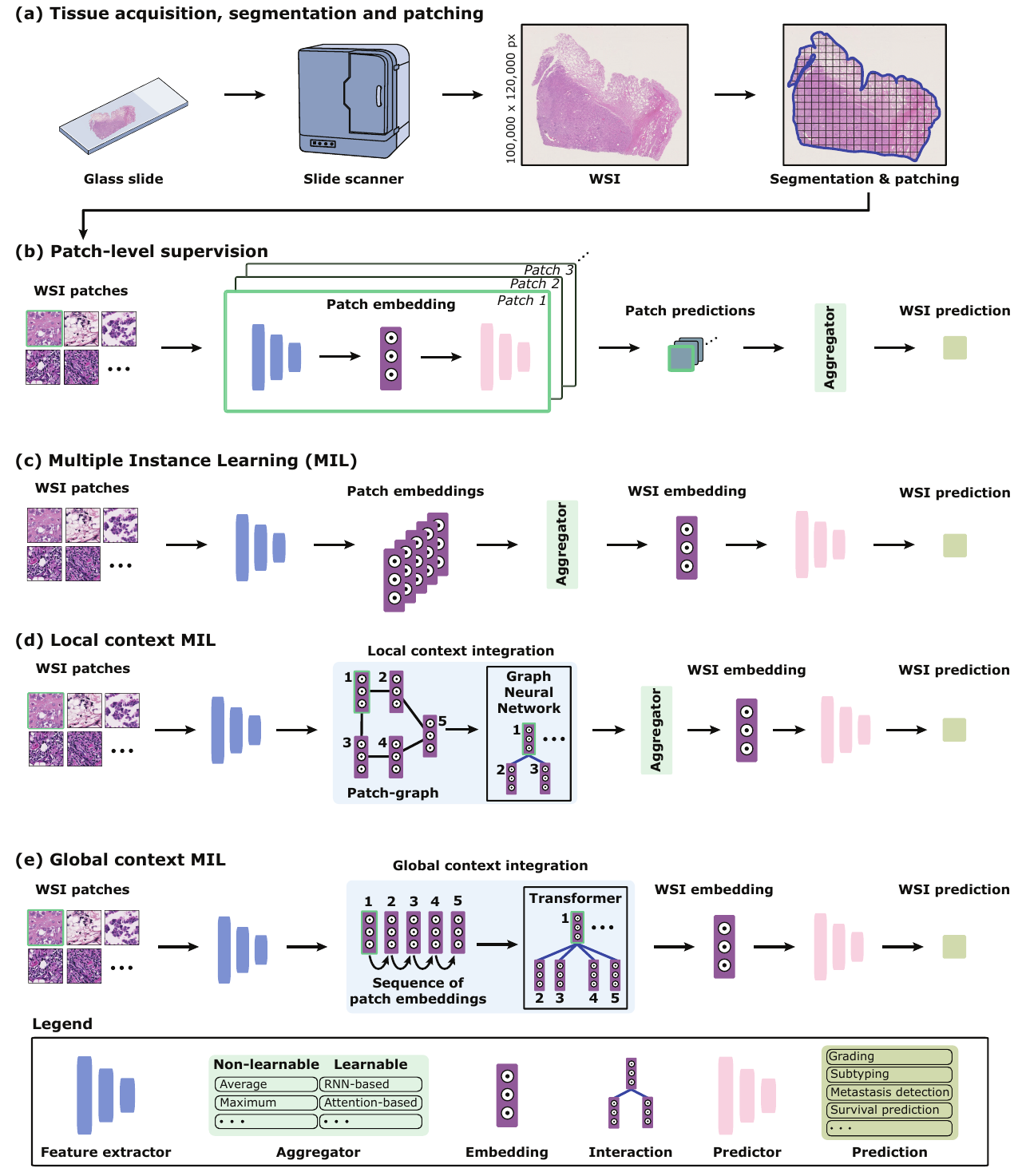}
    \caption{\textbf{Caption next page.}.}
    \label{fig:attention}
\end{figure}
\setcounter{figure}{1}    
\begin{figure}
    \centering
    \caption{\textbf{Multiple instance learning for clinical end-point prediction on whole-slide images.} All methods take as input a set of patches extracted from a whole-slide image (WSI) at a fixed magnification and learn to map it to a WSI-level clinical end point, such as cancer grade or subtype. \textbf{a}, Histology slides are first digitized with a scanner as WSIs, which then go through segmentation and patching. \textbf{b}, For patch-level supervision, each patch is assigned a label, either by using manual patch-level annotations or by assigning the slide-level label to all patches. Patches are passed through a sequence composed of a feature extractor, patch-level predictor and aggregator to produce a WSI-level prediction. \textbf{c}, In multiple instance learning (MIL), a feature extractor extracts embeddings for all patches, which are then aggregated (without including context) for WSI-level prediction. \textbf{d,e}, For context-aware MIL, the interactions between patch embeddings (extracted with the feature extractor) are explicitly encoded using either a graph representation of patches processed with a graph neural network (part d) or a sequence representation of patches processed with a transformer (part e).}
    \label{fig:attention2}
\end{figure}

The methods discussed so far assume that the WSI patches are independent of each other and have no access to contextual information besides what is in the patch. However, this view limits the incorporation of long-range context required to model the tissue architecture\cite{jaume2022integrating}, especially in cases where the local cell morphology alone cannot predict the target (for example, for therapy response prediction)\cite{taube2018implications}. Although using lower magnifications would be a simple workaround, this strategy risks causing the user to miss out on granular cellular details. To address this limitation, MIL methods could, with a dedicated neural network, explicitly model the interactions between patch embeddings. This approach requires the construction of a relational structure of embeddings — either as a graph or a sequence — and the application of a network — either a graph neural network (GNN) or a transformer — that can integrate and model the interactions between embeddings based on the specified structure. Alternatively, MIL methods can implicitly incorporate context by aggregating patch embeddings from multiple magnifications.

\textbf{Graph representations and graph neural networks.} One way of representing interactions between the different patches is to use a graph\cite{pati2022hierarchical,chen2021whole,adnan20}, in which patch embeddings represent nodes that are connected via edges (Fig.~\ref{fig:attention} (c)). Connections (or edges) are generally defined based on a locality principle according to which regions that are physically close to each other are more likely to interact and should therefore be connected\cite{chen2021whole,pati2022hierarchical}. For instance, a patch could be connected to its five closest neighbors or to all its adjacent patches. A graph neural network (GNN)\cite{scarselli2008graph} (a class of neural networks specifically designed to learn on graph-structured data) is then learned to predict the target of interest from the graph. In GNNs, patch representations are passed along edges \andrew{using message passing (i.e., through a series of linear and nonlinear activation operations)} to capture local and global information from the TME jointly. Contextualized node embeddings are then aggregated to form a WSI embedding. A GNN can be trained on a graph with an arbitrary number of nodes and edges, which can be applied to any WSI dataset without constraints on the size and shape of the tissue.

Graph-based MIL approaches are not limited to working at the patch level and can also be used to model nuclei (nodes) and interactions between nuclei (edges), a representation referred to as a cell-graph\cite{pati2022hierarchical,gunduz2004cell,zhou2019cgc,chen2020pathomic}. A cell embedding is often extracted around each nucleus to characterize its appearance. This formulation resembles how biological systems interact, because the nuclear morphology and the cellular interactions are explicitly encoded. A GNN is typically trained to map the cell-graph to a clinical end point, analogous to patch-graph approaches. However, because the number of nuclei in a WSI can be very high (up to several million), scaling to large tissue regions remains challenging\cite{ahmedt21}.

\textbf{Sequence representations and transformers.} Instead of restricting interactions to be based on locality as done with GNNs, one can instead assume that all patches are interacting with each other, regardless of their positions in the image (see Fig.~\ref{fig:attention} (d)). This is the core idea behind transformers\cite{vaswani2017attention,dosovitskiy2020image}. Intuitively, for each patch embedding, a transformer assesses the importance of all other patch embeddings towards contextualizing its own representation, \andrew{a concept known as self-attention. After a series of self-attention operations, the set of patch embeddings gets aggregated into the final global context-aware WSI embedding}. Following the transformer terminology, all the patch embeddings are represented as a sequence, with the position within the sequence indicating the spatial location of the patch in the WSI. Conceptually, this can be seen as a generalization of GNN where all the patches would be connected to each other (i.e., including global context) rather than being restricted to local connections only. However, as the number of interactions is quadratic in the number of patches, this approach has high computational requirements (both in terms of training time and GPU RAM). This limitation makes end-to-end training with the current hardware even more complex. To reduce the computational burden, transformers with lower computational complexity have been proposed by reformulating or approximating the interactions between patches\cite{shao2021transmil,wu2022flowformer}. Broadly, interactions between patches have also been implemented with recurrent neural networks\cite{campanella2019clinical, iizuka2020deep} or variants of transformers\cite{kalra2020learning}.

It remains unclear whether transformer-based approaches are better than graph-based approaches with respect to predictive performance, robustness to domain shifts and generalization capabilities. On the one hand, for applications in which the type of context is known \textit{a priori} (for example, according to some locality principle), graph representations mimic more closely the interaction between biological systems and offer more control over the interactions between patches. On the other hand, transformers have a lower inductive bias, so they impose fewer constraints on the network. Therefore, by learning the graph connectivity with attention weights, transformers are more likely to afford the discovery of new long-range contextual biomarkers.

\textbf{Multiscale representations.} Instead of including context at a single magnification, context can be implemented by using multiscale representations of WSIs, for example, 5$\times$, 10$\times$, and 20$\times$ using a late-fusion method\cite{lipkova2022deep,sirinukunwattana18,skrede2020deep}. This approach aggregates (for example, with concatenation or summation) WSI representations of different magnifications or extracts concentric patches at multiple magnifications\cite{sirinukunwattana18}. It must be emphasized that these strategies are agnostic to the underlying MIL framework used. Alternatively, a dedicated mechanism can be implemented to learn how to zoom into diagnostically relevant regions\cite{thandiackal2022zoommil,katharopoulos19,kong21}, in a similar manner to how a pathologist examines a diagnostic WSI, thereby reducing computation as a result of not having to process all WSI patches at different magnification levels.

\subsubsection*{2.3 AI-based assistive tools} 

Deep learning in pathology has also been used for developing AI-based assistive tools that can extract actionable objects and representations from WSIs for subsequent clinical or research use. These tools have been mainly developed for tissue and nucleus segmentation, as well as for virtual staining.

\textbf{Segmentation.} An essential tool of CPath is to segment WSIs into different components, for example, nuclei, glands or tissue regions. Segmentation is crucial for assisting clinical diagnosis by objectively and quantitatively correlating the morphological traits of clinical outcomes. Segmentation can be either semantic, in which the aim is to assign a morphological class label to each pixel, or it can consist of instance segmentation, which additionally assigns an instance identifier to each object occurrence. Semantic segmentation has mainly been used for epithelium versus stroma segmentation\cite{bulten2019epithelium}, Gleason pattern detection\cite{anklin21} or histological tissue segmentation\cite{chan2019histosegnet}. Instance segmentation is instead used to segment nuclei\cite{graham2019hover}, glands\cite{sirinukunwattana17} and mitotic cells\cite{li2018deepmitosis}, in which the delineation of each entity is important, for example, for measuring HIFs from each cell and gland.

Most deep learning-based segmentation methods operate in a fully supervised setting, and, as such, require fine-grained pixel-level labels, which meticulously detail each entity or tissue in the image. Semantic segmentation networks often use fully convolutional networks\cite{long2015fully} and U-Net\cite{ronneberger2015unet} architectures. Specifically, a CNN encoder compresses the input (typically a tissue image patch) into a spatially aware embedding, in which each element corresponds to a specific location in the original input and a symmetrical CNN decoder expands and converts the embedding into a segmentation mask with a label assigned to each pixel. These networks can be modified to transform the semantic segmentation output into instance segmentation, either with a dedicated additional branch\cite{graham2019hover} or by implementing post-hoc steps. Mask-regional CNN\cite{he2017mask} can also be used for instance segmentation, in which an object detector is first applied to identify objects in the image using a dedicated detection branch operating on latent embeddings. Each detected object is then segmented by directly assigning a label and an instance identifier to each pixel.

Despite deep learning frameworks already achieving impressive performance in segmentation tasks, the annotation process requires substantial resources and support from pathologists. To address this limitation, dedicated annotation tools\cite{kumar17,koohbanani2020nuclick} or the human-in-the-loop approach to interactively correct the predictions of the model\cite{greenwald2022whole} have been introduced. Moreover, weakly supervised semantic segmentation pipelines can segment images from large patches or even WSIs using coarse labels\cite{anklin21,chan2019histosegnet}. However, their performance remains inferior to their fully supervised counterparts. In parallel with H\&E staining, segmentation methods are being applied to immunohistochemistry (IHC) and multiplex images, for example, for nuclei and membrane segmentation\cite{andreou2022multiplexed,greenwald2022whole,martinelli2022athena,han2022cell}.

\textbf{Virtual staining.} Virtual staining consists of transforming the appearance of an image with an algorithm. Two applications exist in CPath: stain enhancement for correcting, normalizing and augmenting stains, and stain transfer for converting the image from one stain/image modality into another. Because of differences in tissue processing and digitizing protocols across institutions, WSIs often have different appearances, which can negatively impact the performance of deep learning systems. Stain normalization can be used to mitigate these biases and increase model performance and robustness to domain shifts\cite{tellez18,zanjani2018stain}. Although stain normalization has traditionally dealt with stain-vector estimation\cite{macenko09,vahadane16}, by mathematically modelling pixel-to-pixel colour mapping, deep generative models can also be used to train a model to generate an input image with the staining intensity of a reference dataset\cite{}. Deep generative models can also convert H\&E frozen sections (generated via a fast procedure that may produce artefacts) into H\&E formalin-fixed paraffin-embedded sections (a slower procedure that is less affected by noise)\cite{cho2017neural, zhou2019enhanced, kang2021stainnet} — an application that provides access to high-quality sections in a shorter time, for example, in surgical operations. Therefore, staining enhancement applications can enhance the reliability of deep learning models and help pathologists by reducing visual variability between samples.

Stain transfer with deep generative models is also a promising direction to transform images from one staining or image modality into another, a particularly difficult task owing to the inherently different imaging protocols between the two modalities. For example, images can be transferred from H\&E staining to IHC and multiplex images\cite{he2022ai,ghahremani2022deep}, and from ultraviolet microscopy to H\&E\cite{cao2022label}. Although these methods are best trained on pairs of registered images in both domains, recently developed algorithms have enabled the use of unpaired data\cite{zhu2017unpaired, park2020contrastive}, leading to a substantial simplification of data collection. Despite no consensus existing on whether virtually stained images are clinically applicable\cite{VASILJEVIC2022cycle}, we believe that, with rapid advances in deep generative models, these images will have an increasing role in CPath.

\subsubsection*{2.4 Interpretable CPath}

\begin{figure}
    \centering
    \includegraphics[width=0.95\textwidth]{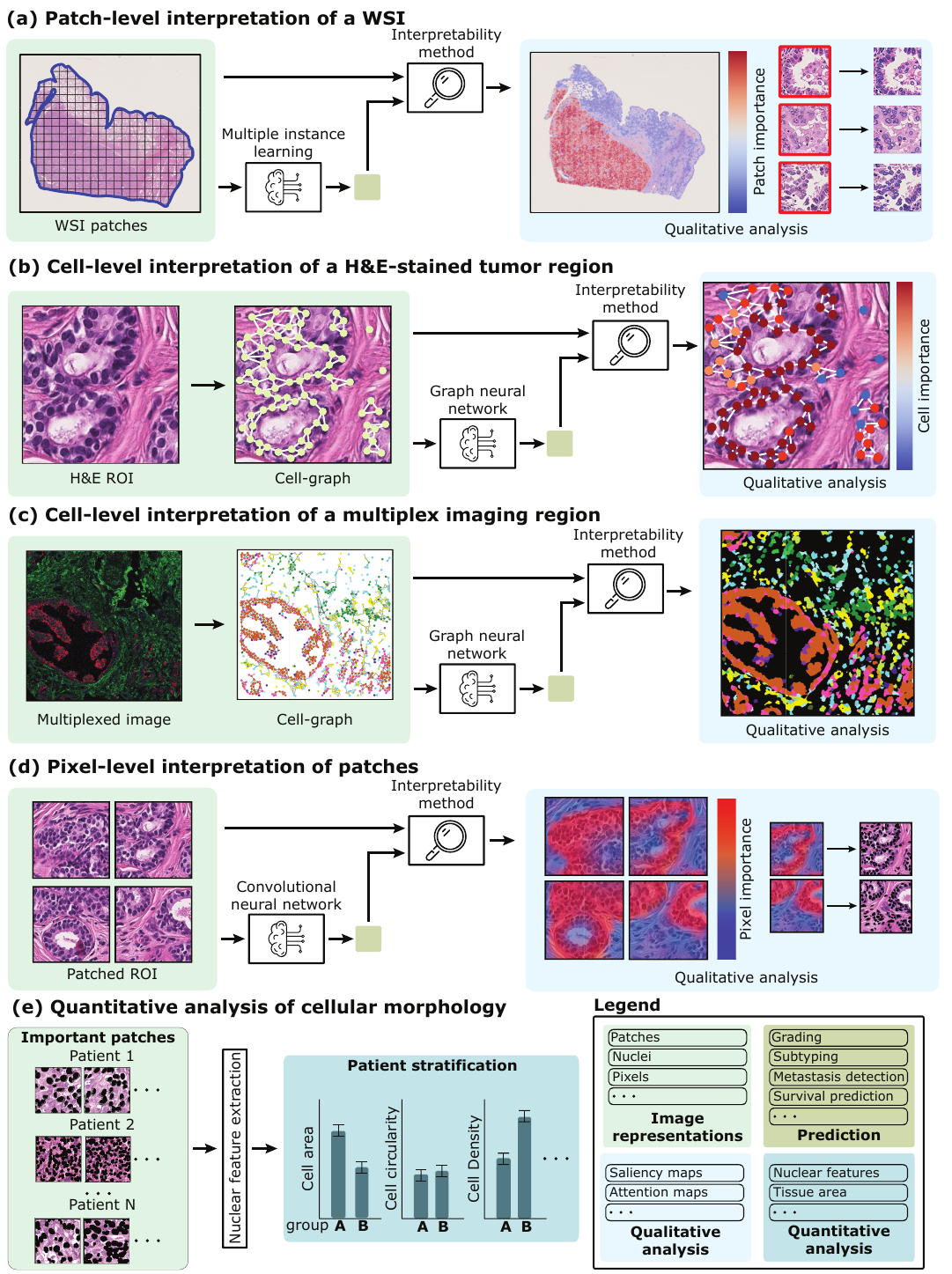}
    \caption{\textbf{Caption next page.}}
    \label{fig:interpretability}
\end{figure}
\setcounter{figure}{2}    
\begin{figure}
    \centering
    \caption{\textbf{Interpretability methods in computational pathology.} A whole-slide image (WSI) is transformed into a representation of interest, for example, a patch-graph or a cell-graph, and fed to a predictive model — a non-interpretable deep learning model — for clinical end-point prediction. On the basis of the prediction, qualitative (for instance, a saliency map) and quantitative (for instance, a tumour cell count) interpretability analyses are performed. The choice of WSI representation dictates the resolution of interpretability output, for example, patch (cell) inputs result in patch-level (cellular-level) importance scores. \textbf{a}, Pipelines based on patch-level multiple instance learning (MIL) can use a qualitative interpretability method, such as feature attribution, for constructing a saliency heatmap. \textbf{b,c}, A cell-graph representation results in nucleus-level importance scores. \textbf{d}, With a patch-level label, pixel-level importance scores can be computed. \textbf{e}, The qualitative analysis outcome is used for a quantitative study of the importance scores to derive WSI or cohort-level insights about the model behaviour. H\&E, hematoxylin and eosin; ROI, region of interest.}
    \label{fig:interpretability2}
\end{figure}

The ability to explain, justify and understand the decisions made by deep learning methods is essential to establish a relationship of trust between AI systems and pathologists\cite{holzinger17} (Fig.~\ref{fig:interpretability}). In particular, the interpretability of deep learning methods requires the identification of important regions that provide insight into the prediction. In clinical settings, these regions can then be used to ensure the reliability of the method by comparing it with expert knowledge and serve as an indicator for the automatic selection of regions of interest. In a research setting, the delineation of morphological features characterizing salient regions can contribute to biomarker discovery. However, owing to the intrinsic complexity of the model, identifying these regions and understanding the mechanism by which deep learning methods make a decision is not always straightforward.

To overcome this limitation, feature attribution methods\cite{selvaraju2017gradcam,sundararajan2017axiomatic}, e.g., based on gradient importance, have been proposed. The central idea is to iteratively recover the deep features that have been most activated by the neural network back to the input (e.g., up to the patch embeddings), thereby resulting in a saliency map where each element of the input is assigned to a score that indicates how important the element is for the prediction. Alternatively, attention-based methods can directly interpret the learned attention weights as importance scores for explaining the prediction\cite{ilse18attention, lu2021ai,lu2021data, chen2022pan, lee2022derivation}. Although both methods have primarily been applied to WSI classification systems, resulting in patch-level importance scores, they can also work with patch classification (resulting in pixel-level scores) or cell-graph classification (resulting in nucleus-level scores).

Despite these analyses providing valuable insights, high importance scores do not necessarily prove the presence of a certain class, and researchers must qualitatively determine the reason for the selection of each region. As a potential solution, reformulation of MIL to derive patch-level predictions from WSI-level label training has been proposed\cite{Javed2022additive}. Another limitation is that these methods are constrained to qualitative considerations; therefore, saliency maps with patch-level resolution cannot readily delineate the morphological features and nuclei responsible for a prediction. Consequently, quantitative morphological characterizations, based on segmentation frameworks and HIFs derived from segmented entities, are required to complement the interpretability analysis to ascertain findings\cite{madabhushi2016image}. These characterizations can be achieved by studying glandular or nuclear morphological descriptors based on shape, size or chromaticity, and graph-based topological descriptors based on density, dispersion or tissue architecture\cite{diao2021human}.

Ultimately, the interpretation of a clinical prediction can include qualitative and quantitative considerations resulting from the combination of saliency maps produced by deep learning-based prediction methods and HIFs. A typical pipeline would first have a deep learning-based predictive model trained for a certain predictive task, for example, for cancer grading. Next, saliency maps for test tissue images would be produced from the predictions using an interpretability tool (such as attention-based saliency map), providing the grounds for qualitative analysis. Finally, HIFs in important regions of the saliency map would be computed and aggregated for quantitative morphological analysis within each WSI or cohort. Examples of this workflow include studying nucleus-level properties and interactions in important patches\cite{lu2021ai,chen2022pan}, as in  Fig.~\ref{fig:interpretability} (a,e), or in important nuclei of cell-graph representations\cite{jaume21}, as in Fig.~\ref{fig:interpretability} (b,e). This general workflow is neither restricted to specific staining protocols, as in Fig.~\ref{fig:interpretability} (c,e), nor specific feature attribution methods, as in Fig.~\ref{fig:interpretability} (d,e).

\section{Public datasets and open-sourced codes}\label{sec:public}

The aforementioned methodological advances have greatly benefited from initiatives to create public datasets, either in the form of challenges or open data banks. Challenges are often proposed as competitions on a well-defined task for which new methodological contributions are needed to advance the field. Popular challenges have targeted mitosis detection in breast cancer\cite{ludovic2013mitosis,veta2019breast,aubreville2022mitosis}, the detection of breast metastasis in lymph nodes (CAMELYON16, CAMELYON17)\cite{bandi2018detection}, Gleason grading of prostate biopsies (PANDA)\cite{bulten2022artificial}, breast cancer subtyping (BACH)\cite{bach18}, pan-cancer nucleus segmentation and classification\cite{monusac2020}, gland segmentation in colon images\cite{sirinukunwattana17}, among others. In CPath, several of these challenges have become reference datasets used to benchmark new methods. The scale of these challenges keeps increasing as the field advances, e.g., from 400 WSIs in CAMELYON17 to more than 10,000 WSIs in PANDA, in just four years. 

In parallel, open data banks, where hospitals can add anonymized clinical data, have been developed. As of today, the main resource remains The Cancer Genome Atlas Program (TCGA), which includes more than 20,000 primary cancer cases spanning 33 cancer types with imaging, omics data, and patient information. 
TCGA has been the main driver for defining new problem statements in CPath. Other initiatives include the Clinical Proteomic Tumor Analysis Consortium (CPTAC), and the National Lung Screening Trial (NLST). Ongoing efforts aiming to build larger banks of pathology data include Big Picture\cite{moulin_imibigpicture_2021}, PathLake, UK Genomics Pathology Imaging Collection\cite{jennings_bridging_2022}, and the UK Biobank. 

To enforce reproducibility and reduce boilerplate code in CPath, several open-source libraries and software programs have been developed to ease the understanding of new publications and to speed up the development of new methods\cite{Wagner2022Make}. Today, libraries and software can be used for efficient reading of WSIs\cite{openslide20}, data visualization and annotation\cite{qupath21}, patching and deep feature extraction on WSIs\cite{lu2021data,histomicstk21}, detection of tissue vs. background regions\cite{lu2021data,histomicstk21}, stain normalization\cite{histomicstk21}, graph modeling\cite{jaume2021}, and multimodal data inputs\cite{rosenthal2022building}. For CPath models, efforts in publicly releasing the code and the trained model checkpoints, such as semantic and instance segmentation networks\cite{Pocock2022}, attention networks\cite{lu2021data}, and pre-trained image encoders on histology images\cite{ciga2022self, chen2022scaling, wang2022transformer}, are also becoming the field standard. Because open-source libraries have an inherent risk of not being maintained on a regular basis by their developers, a concerted community-level effort to support them remains necessary in the future.

\section{Clinical impact of CPath}

\begin{figure}[h]
    \centering
    \includegraphics[width=\textwidth]{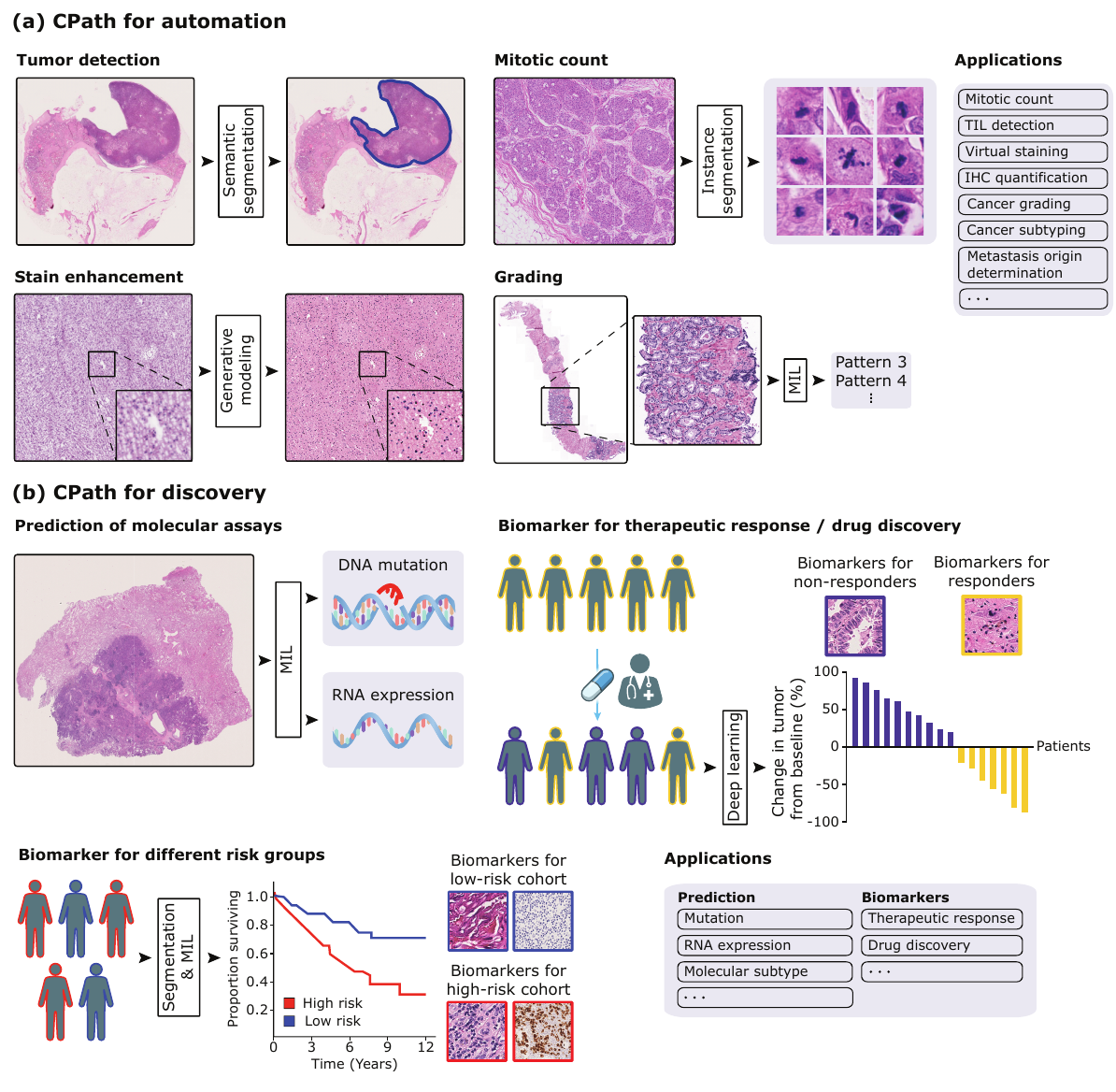}
    \caption{\textbf{Integration of computational pathology in pathology.} Tasks in computational pathology (CPath) can broadly be categorized into automating existing pathology routines and guiding biomedical research. Both categories rely on whole-slide classification systems, artificial intelligence-based assistive tools, such as segmentation networks, or a combination of both. \textbf{a}, CPath for automation replaces time-consuming manual work that pathologists already perform, from mitotic counts to cancer subtyping tasks, thereby alleviating the burden on pathologists and decreasing interobserver variability. \textbf{b}, Outside the clinical practice, CPath frameworks are used for biomedical research, for example, for the discovery of morphological correlates of molecular alterations and different risk/response groups. MIL, multiple instance learning.}
    \label{fig:workflow}
\end{figure}

Since its inception, CPath has pursued two primary and non-mutually exclusive goals: automating portions of the routine clinical workflow and gaining new insights using data from that same workflow, sometimes supplemented by additional data sources (Fig.~\ref{fig:workflow}).

\begin{figure}[h]
    \centering
    \includegraphics[width=\textwidth]{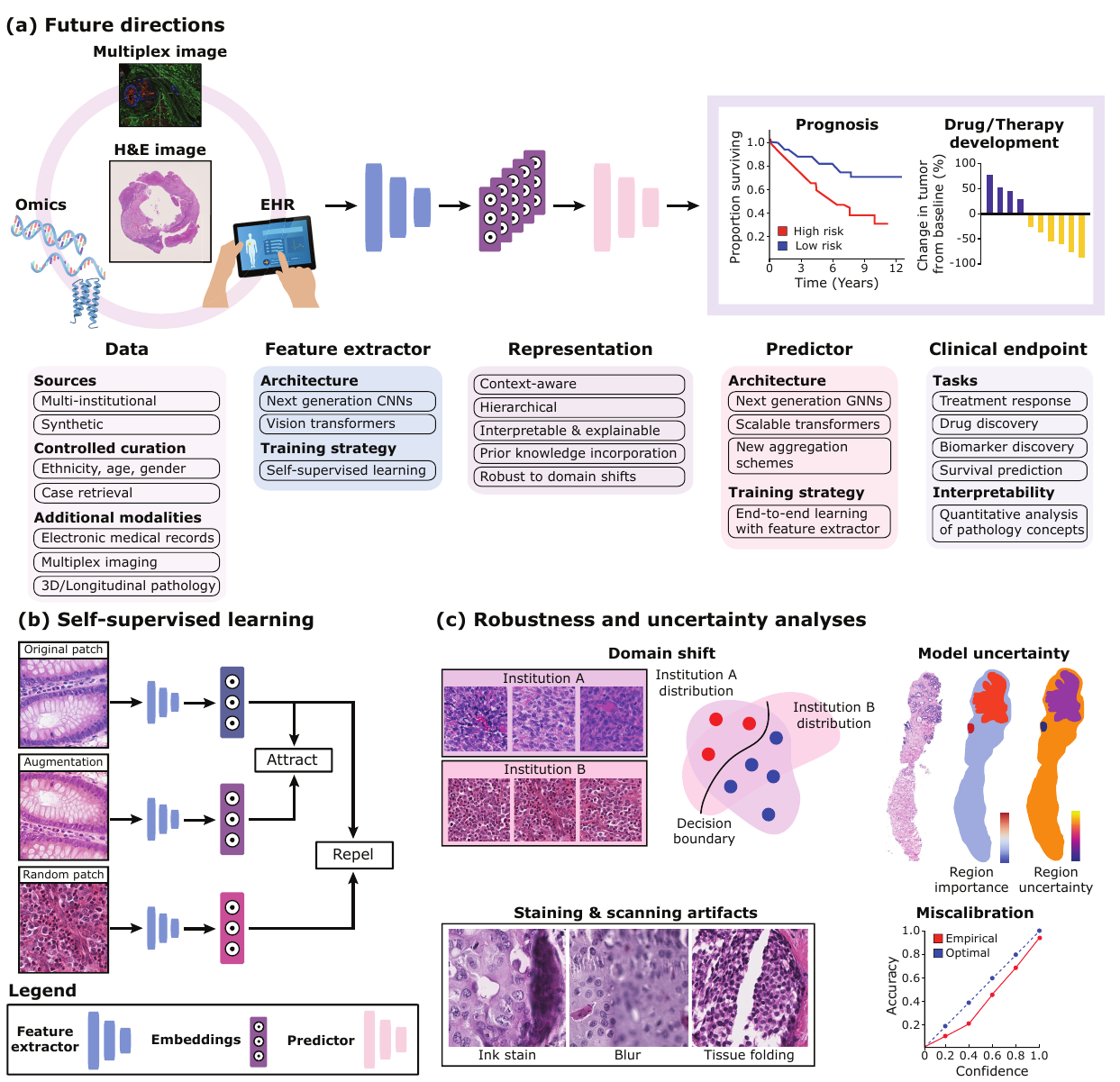}
    \caption{\textbf{Future directions in computational pathology.} \textbf{a}, Computational methods depend on data quality and availability. The development of new, curated, multi-institutional and multimodal cohorts is essential to accelerate the development of computational pathology (CPath), which is expected to evolve further into tasks such as prognostic prediction and biomarker and drug discovery. \textbf{b}, New training strategies, such as those based on self-supervised learning, are needed to build more generalizable representations of histopathological data. \textbf{c}, Accounting for uncertainty will become increasingly important to ensure robustness to domain shifts (for example, owing to image preparation artefacts) and model calibration. CNN, convolutional neural network; EHR, electronic health record; GNN, graph neural network; H\&E, haematoxylin and eosin.}
    \label{fig:outlook}
\end{figure}

\subsubsection*{4.1 CPath for automation}

CPath for automation seeks to recapitulate and augment morphologically well-defined tasks that pathologists already perform, implicitly or explicitly, during their day-to-day work.

\noindent \textbf{Cellular and tissue levels.} Tasks that are tedious and subjective to significant interobserver variabilities such as mitosis detection and counting\cite{cirecsan2013mitosis, li2018deepmitosis, tellez2018whole,veta2019breast,aubreville2022mitosis} and quantification of IHC\cite{kapil2018deep, swiderska2019learning, fassler2020deep, ghahremani2022deep} can be especially attractive targets for researchers wishing to improve the practice of pathology. Instance segmentation of structures, especially nuclei, is a common strategy for automation\cite{graham2019hover,naylor2019segmentation}. An enormous variety of downstream analyses can be performed using the resulting output, such as cell-graph modeling for tissue grading and subtyping\cite{gunduz2004cell,chen2020pathomic,pati2022hierarchical,jaume21}, \andrew{automatically quantifying percentage of programmed death-ligand (PD-L1) positive tumor cells\cite{widmaier2020comparison}}, or the detection of malignant white blood cells\cite{matek2019human}.

Moving up a level in the biological hierarchy, the \andrew{semantic} segmentation of tissue structures have been another popular application of CPath\cite{chan19histosegnet}, including epithelium and gland segmentation\cite{binder19,bulten2019epithelium,graham2021lizard}, vessel and nerve segmentation\cite{fraz2020fabnet}, and prostatic adenocarcinoma gland segmentation\cite{ing2018semantic}. As in the case of cell segmentation, segmented tissues are often combined to form TME-level features\cite{diao2021human} and prognostic biomarkers, such as tumor-stroma area ratio\cite{geessink2019computer} or tumor-infiltrating lymphocyte (TIL) assessment\cite{amgad2020report}. \andrew{Both the cellular and tissue-level segmented results could also be simultaneously considered. For instance, a histologic grading system for Non-alcoholic fatty liver disease (NAFLD) and nonalcoholic steatohepatitis (NASH) integrates features from the cellular level (liver cell injury and lobular inflammation) as well as the tissue level (fibrosis), which benefit from the segmentation frameworks\cite{taylor2021machine, heinemann2022deep}.}

\noindent \textbf{Whole slide level.} AI-based classification of WSIs represents a very active field of study, offering substantial prospects for improving diagnosis and its reproducibility. These algorithms integrate information from an entire slide, not just a single cell or region of interest, in a similar manner to how pathologists must examine an entire slide to formulate a final diagnosis. These approaches can reduce the substantial interobserver variability for some tasks, which, for example, is nearly 50\% for atypia detection in breast cancer\cite{jaume2021} and is inevitable despite multiple years of specialized training.

Grading, defined by the appearance of abnormal cells and tissues relative to their healthy counterparts, which is quantified by pathologists according to features such as gland morphology and nuclear pleomorphism, is an essential aspect of the diagnosis of many diseases, including cancer. In particular, Gleason grading of prostate cancer has been an area of intense focus in CPath owing to the large volume of prostate cancer biopsies in many clinical practices and the high interobserver variability of the task. Multi-institutional studies analysing thousands of samples can now achieve Gleason grading performance that is on par with or exceeds that of pathologists\cite{strom2020artificial, bulten2020automated,bulten2022artificial}. AI-assisted grading has also been applied to other types of cancer, including gliomas\cite{ertosun2015automated}, colorectal carcinoma\cite{zhou2019cgc}, breast cancer\cite{couture2018image}, among others, as well as in heart\cite{lipkova2022deep} and kidney\cite{kers2022deep} allograft rejection. 

Diagnosis in CPath is also often formulated as a subtyping problem, in which an algorithm aims to classify cases within a group of diagnoses; for example, classifying lymph nodes based on the binary presence or absence of metastases or identifying cases of non-smallcell lung cancer as adenocarcinoma or squamous cell carcinoma. These tasks are often presented as multiclass classification problems in which both patch-level supervision and MIL can be applied. Subtyping has been studied in a variety of diseases, including colorectal cancer\cite{korbar2017deep}, skin cancer\cite{ianni2020tailored}, gastric and colon cancers\cite{iizuka2020deep}, liver cancer\cite{kiani2020impact}, breast cancer\cite{pati2022hierarchical,bach18}, lung cancer\cite{shao2021transmil, lu2021data}, and the detection of lymph node metastases\cite{bejnordi2017diagnostic,bandi2018detection,zhao20}, even predicting whether a tumour is primary or metastatic and identifying its primary site, if metastatic\cite{lu2021ai}.

Applications built on the aforementioned techniques have the power to truly augment clinical practice in significant ways, especially in combination with AI models that can convert the appearance of the tissue to the domain familiar to clinicians\cite{kutsev2021deep, cao2022label} or utilize alternative imaging techniques\cite{hollon2020near}. Applications such as an AI-assisted augmented reality microscope\cite{chen2019augmented} and AI-based triaging tool to reduce pathologist burden\cite{gehrung2021triage} illustrate how clinical practice can be further enhanced. \andrew{Finally, CPath tools can serve as prescreening tools and reduce the number of follow-up tests, thereby reducing turnaround time to reach the final diagnosis. For instance, the prediction of top primary candidate sites for metastatic cancer can reduce the number of required IHC tests\cite{lu2021ai}. Differently, the microsatellite stability makers can be identified with high sensitivity, thereby helping prescreen colorectal cancer patients who do not require further microsatellite instability (MSI) testing\cite{kather2019deep}.}

\subsubsection*{4.2 CPath for discovery}

Prognostication, molecular marker prediction and biomarker discovery use histology, genomics and other data modalities to perform tasks that pathologists — and treating clinicians — do not, either because they were never trained to do so or because these tasks mainly rely on the ability of deep learning to identify connections in high-dimensional data, an ability that humans do not possess.

\noindent \textbf{Prognosis.} Prognostic models aim to predict the risk of disease progression, for example, by assessing the risk of cancer recurrence and metastasis, and, ultimately, the probability of patient survival. Prognosis prediction is particularly difficult because morphological correlates of prognosis are generally not well understood. This can be explained by interpatient heterogeneity, which creates ambiguity and hinders the establishment of one-fits-all criteria; moreover, prognosis is the combination of multiple factors, such as genetic markers, clinical traits such as age or comorbidities, and treatment response, only some of which might be reflected in the tissue. Furthermore, the limited understanding of morphological correlates for prognosis renders manual annotation of tissue challenging, making approaches that do not rely on manual pixel annotations, such as MIL frameworks, attractive.

Owing to the complexity of prognosis, patients are often stratified into risk groups with substantially different outcomes rather than being individually assigned an exact predicted survival. Several pan-cancer and cancer-specific studies suggest that histological data can be pre- predictive of patient survival\cite{bychkov2018deep,kather2019predicting,skrede2020deep, chen2022pan}, recurrence risk\cite{leo2021computer, wang2017prediction, shaban2019novel, kulkarni2020deep}, and risk of metastasis\cite{yang2022prediction, klimov2021predicting}. In particular, some survival analyses have shown that DL-based patient stratification presents better separation between risk groups than existing well-known, prognostic biomarkers\cite{skrede2020deep, lee2022derivation}, such as clinical grade and stage. These advances have the potential to enhance patient treatment and can help provide better-tailored therapies to each sub-cohort\cite{kleppe2022clinical}. Additionally, the comparison of histologic features and tissue composition of highly-attended regions of different risk groups can further reveal prognostic biomarkers unbeknownst to pathologists\cite{courtiol2019deep, saillard2020predicting}, which could potentially guide target identification for therapy development. For instance, in a rigorously-validated colorectal cancer survival study\cite{skrede2020deep}, a DL-based prognostic biomarker was able to stratify stage II and III patients into more fine-grained and distinct prognostic groups. This can guide the assignment of tailored adjuvant treatment regimes, potentially leading to better patient outcomes compared to treating patients based on traditional stages. Recent works emphasize the importance of long-range dependencies for prognosis prediction to better capture tumor heterogeneity\cite{jaume_song_mahmood_2022,zhao2020predicting, chen2021whole,lee2022derivation} by relying on graph representations and GNN. While not clinical-grade, performance is expected to increase with larger cohorts and less noisy prognostic endpoints.

Given that disease progression and therapy response are mainly governed by complex processes involving different genotypic and phenotypic factors, prognostic research has increasingly included context within WSIs\cite{boehm2022harnessing, roelofsen2022multimodal} (for instance, by relying on graph representations) or using additional modalities\cite{boehm2022multimodal,roelofsen2022multimodal}. Specifically, multimodal integration has involved histological (for example, patch embeddings and cell features) and radiological (such as radiomics features) data, as well as the results of molecular assays (such as copy number variation, mutation status and bulk RNA sequencing) and the evaluation of clinical variables (such as age and sex)\cite{cheerla2019deep,chen2020pathomic,chen2022pan}. In a pan-cancer study\cite{chen2022pan}, for example, combining histopathological data and molecular assay results improved the reliability of survival prediction in most of the investigated cancer types, compared with using either modality alone. Similarly, combining IHC stains of different immune cells improves prediction of risk of colorectal cancer relapse\cite{foersch2023multistain}.

\noindent\textbf{Molecular profile prediction.} The connection between histological and molecular assay data, such as IHC and bulk DNA sequencing data, is also being actively explored. Although the cost and complexity of these assays continue to decrease, they are still expensive, time-consuming and not routinely conducted. Therefore, models that can accurately predict these results, whenever possible, from the relatively cheap and widely available H\&E images, are attractive. Such models could also contribute to discovering new morphological biomarkers that correlate with molecular alterations. Identifying these morphological traits is important for developing new drugs and therapies as they reveal how targeted treatments affect tissue morphology\cite{vamathevan2019applications}. Several works have shown that molecular alterations affect cellular morphology and the surrounding TME in specific ways\cite{echle2021deep}. For instance, histology slides can predict certain mutations, as shown in pan-cancer\cite{kather2020pan, fu2020pan} and disease-specific studies\cite{wang2019predicting, loeffler2022artificial}. Histology has also been shown to be predictive of gene expression\cite{schmauch2020deep}, MSI\cite{kather2019deep}, molecular subtypes\cite{hong2021predicting}, PD-L1 status\cite{shamai2022deep}, and protein expression\cite{naik2020deep}. Novel biomarkers can be extracted by studying the morphology of regions that are responsible for prediction, thereby providing a better understanding  of the interactions between phenotype and genotype. Most recent studies use slide-level molecular assay labels, a limitation of common bulk sequencing assays that can only provide a WSI-level description, with both the patch-level supervision with assigned slide-level signatures approach and the MIL approach used for learning. Recently, the increasing availability of spatially-resolved sequencing technologies such as spatial transcriptomics has ignited interest in understanding the connections between specific histomorphologies and single-cell molecular profiles. Successful prediction of spatial transcriptomic data\cite{he2020integrating} and spatially-resolved protein expression via IHC\cite{acosta2022intratumoral, song2022investigating} from H\&E images opens up avenues for further studies in this area.

\noindent\textbf{Therapeutic response prediction.} CPath could also assist in predicting patient response to treatment. For example, AI systems have been developed to predict response to immunotherapies\cite{harder2019automatic, hu2021using,berry2021analysis}, targeted therapy\cite{farahmand2022deep,bychkov2021deep}, and chemotherapy\cite{li2021deep}. Because the assessment of therapeutic response relies on different patient signatures, it is not surprising that many of these studies are multimodal in nature\cite{sammut2022multi, vanguri2022multimodal}. However, the difficulty in collecting large datasets remains a major challenge, often requiring retrospective analyses of clinical trial data. Instead, models for mutation prediction from histology can be used as a surrogate, as high tumour mutational burden and microsatellite instability might themselves be predictive of response to therapies\cite{liu2022systematic}.

\section{Outlook and future directions}

Despite remarkable advances in CPath, challenges and opportunities remain, which can be grouped into advances related to data acquisition and processing, those related to building new AI methods\cite{bera2019artificial, niazi2019digital, kleppe2021designing, van2021deep, baxi2022digital} (Fig.~\ref{fig:outlook}) and translational considerations for clinical deployment. 

\subsubsection*{5.1 Data outlook}

Reaching the goal of precision medicine requires that increasingly different assays are carried out, each with a finer spatial and temporal resolution, thereby drastically increasing the amount of data available per patient. Not only does this trend complicate data collection but it also presents new methodological challenges.

\textbf{Case retrieval.} Most practices have a searchable database of pathology reports, which are helpful for many tasks, from finding how a patient’s previous pathology was signed out, to identifying cases of rare diseases, to later examining their morphologies. These textual databases contain only a tiny fraction of the information on the slides that led to the creation of those pathology reports, especially in practices that do not include microscopic descriptions. Tools that would allow pathologists to search the images themselves unlock a host of other possibilities\cite{kalra2020yottixel, chen2022fast}, such as simplification of the finding of cases that share a certain morphology to help make a diagnosis, detection of misidentified cases and comparison of the primary tumour morphology of a patient to assess potential risk of metastasis.

\textbf{Multimodal integration.} Despite proven improvements in prognostic performance when combining data from multiple modalities, challenges remain before their full potential can be realized. First, existing multimodal datasets are rarely larger than a few hundred samples (as is the case of the omics and histology TCGA cohorts), which is much smaller than CPath studies with tens of thousands of WSIs\cite{campanella2019clinical,lu2021ai}. However, the transition to larger datasets constitutes a major logistical challenge; in retrospective studies, multimodal data might be scattered across multiple institutions, complicating data collection and heightening the risk of missing modalities for many patients. In prospective studies, the overall cost can increase dramatically with the number of modalities; for example, whole-genome sequencing, as in TCGA, remains prohibitive for most institutions. Even with sufficient financial resources, amassing these data in sufficient quantities can take years. Nevertheless, given the excitement around multimodal studies, we expect more work to be done combining histology, radiology, electronic medical records and next-generation imaging, such as spatially resolved transcriptomics. The prospect of larger datasets with more modalities has already led to new methodological advances; for example, early-fusion methods are now being explored to account for local cross-modal interactions directly within the encoding pipeline\cite{chen2021multimodal}. Moreover, methods for handling missing modalities (without trivially removing cases with missing modalities) are also gaining traction, for example, with multimodal dropout\cite{cheerla2019deep,lipkova2022deep}.

\textbf{Encoding temporality.} Longitudinal (or temporal) study of disease progression is crucial for continuously monitoring patient health, for example, to understand the dynamics of heart and kidney transplant rejection or to help determine the effectiveness of new cancer treatments in clinical trials. However, longitudinal data collection is logistically complex, because assays are not always conducted by the same institution or acquired at consistent intervals within and between patients. The next iteration of data initiatives, such as The Cancer Moonshot initiative\cite{singer2022new}, can transform this picture with the large-scale collection of longitudinally tracked cancer data. Because longitudinal tissue biopsies cannot be extracted at the exact same anatomical location, modelling and integrating intra-organ heterogeneity will be particularly challenging and require new methodological contributions.

\textbf{3D pathology.} Human tissue is by nature 3D, whereas CPath analyses are mainly based on 2D sections, thus risking overlooking important morphological regions\cite{liu2021harnessing}. To overcome this limitation, one option is to perform 3D reconstruction from serially sectioned and aligned H\&E tissue sections to visualize the entire tissue at subcellular resolution, without the need for special sample preparation protocols or equipment\cite{kiemen2022coda}. Alternatively, advanced imaging techniques, such as open-top light-sheet microscopy\cite{glaser2017light}, microscopy with ultraviolet surface excitation\cite{fereidouni2017microscopy} and micro-computed tomography\cite{katsamenis2019x}, can be used to capture high-resolution tissue-preserving 3D representations of tumours. This approach avoids the destructive microtome slicing process for preparing 2D slices, the implementation of which could alter the original tissue morphology. With the promise of better characterization of disease with 3D morphology and more accurate patient prognosis\cite{xie2022prostate}, we anticipate 3D CPath research to progress, especially with computer vision methods for 3D modelling that could be applied to 3D pathology becoming more available. However, challenges remain as imaging tools remain prohibitively expensive for clinical deployment, and current pathology education curricula are solely based on 2D morphology. At least in the foreseeable future, these techniques are likely to be used for research purposes before they can be integrated into routine clinical workflows.

\textbf{Next-generation imaging.} Emerging multiplex imaging techniques\cite{andreou2022multiplexed}, such as immunofluorescence and chromogenic IHC, and mass or flow cytometry, allow the simultaneous assessment of multiple biomarkers in the same tissue section, down to single-cell resolution. Each technique has advantages and limitations in terms of a number of markers, resolution, maximum image size, acquisition time, and cost. Multiplex imaging is particularly promising for understanding tumour heterogeneity in the tumour microenvironment and discovering cellular and molecular features predictive of treatment response — tasks that cannot necessarily be carried out from H\&E images alone\cite{allam2020multiplex}. In CPath, multiplex image studies are mostly based on instance segmentation algorithms that enable nucleus-level analysis, using either handcrafted features\cite{berry2021analysis,martinelli2022athena} or GNNs on cell-graphs\cite{wu2022graph}. As the size of multiplex image cohorts increases, we expect that the recipes that have made H\&E image analysis successful, for example, with MIL, will be adapted to multiplex images\cite{lu2021multiplex}.

\textbf{Decentralized learning.} Although collecting large-scale cohorts through multi-institutional collaborations offers the prospect of more diverse data, the cost of data storage, privacy concerns and the risk of compromising competitive edges often mean that data cannot be stored in a single location\cite{price2019privacy, shmatko2022artificial}. This limitation demands the development of new decentralized training strategies — a task known to be challenging for deep learning models. In particular, asynchronously populating updated model parameters across different locations during multi-institutional training is not trivial. Different techniques to mitigate the difficulty of implementing decentralized training have been proposed, such as federated\cite{zhang2022shifting,lu2022federated, adnan2022federated}, continual\cite{parisi2019continual} and swarm\cite{saldanha2022swarm} learning; however, despite the encouraging results, these techniques remain proofs of concept and are yet to be adopted by clinical institutions. Even when multi-institutional collaborations are established, ensuring and maintaining a consistent software stack between institutions whose information technology infrastructure was not designed to train large-scale deep learning systems are major challenges. Nonetheless, we expect decentralized learning to be translated into large-scale studies that could not have been conducted otherwise, for example, for modelling rare diseases for which each institution has only a few patient-collected samples.

\textbf{Fairness and biases.} Algorithmic biases that result from disparities in race, socioeconomic status, and gender in datasets remain poorly understood. Seminal works have identified such biases in existing models, for instance, when estimating patient risk\cite{obermeyer2019dissecting, seyyed2021underdiagnosis}. In CPath, most datasets are biased toward individuals of European descent (e.g., in TCGA, 82.0\% of cases are from Whites, 10.1\% from Blacks or African Americans, and 7.5\% from Asians). Therefore, it is crucial to understand the consequences of these biases and to ensure that DL methods work equally well for different groups\cite{ricci2022addressing}. Biases can be, to some extent, mitigated with the acquisition of more diverse datasets better encompassing phenotypic diversity and ensuring representation of all minorities. Building such a large, unbiased cohort is therefore an essential step for the future of CPath, which will need to rely on multi-institutional collaborations. Along with better data collection, synthetic data generation\cite{chen2021synthetic, zhang2022shifting} with deep generative models can be used to increase the proportion of cases from underrepresented groups. In addition, new training methods can be developed for mitigating biases, for instance, based on weighted sampling\cite{krasanakis2018adaptive, jiang2020identifying}. 

\subsubsection*{5.2 Learning better representations}

Developing tools for learning better representations of pathology data is also of crucial importance, particularly for building robust and generalizable WSI representations.

\textbf{Feature extractor architecture.} Neural networks used for extracting patch embeddings have largely followed trends from computer vision. From the use of CNNs with large convolutional kernels (such as AlexNet), the community has adopted deeper CNNs with smaller kernels and residual connections (such as ResidualNet and EfficientNet) and is now exploring vision transformers\cite{dosovitskiy2020image,chen2022scaling} or hybrid CNN–transformer architectures\cite{wang2022transformer}. We expect that developments in both CNNs and vision transformers will keep being adopted for CPath applications.

\textbf{Self-supervised training (SSL) for histopathology data.} SL is a promising training paradigm for representation learning — in particular for building patch embeddings. Instead of using external labels, SSL extracts a training signal directly from the training data by exploiting its structure, hence the term self-supervised\cite{krishnan2022self}. Although, in theory, training a model end to end (from the feature extractor to the predictor) could provide the best predictive performance, there is a risk of overfitting, and thus a lack of generalizability. Transfer learning based on SSL pre-trained feature extractors with histopathological data is particularly promising as an approach to addressing this limitation, because SSL can derive domain-specific patch embeddings without the need for annotations.

Numerous patch-level SSL training strategies have been proposed in CPath\cite{li2021dual, huang2021integration,ciga2022self}; for example, one can learn to map a patch and an augmented version of it (typically based on histopathology-relevant transforms such as random rotation, crop and stain jitter) to similar embeddings (Fig.~\ref{fig:outlook} (b)), learn the correspondence between global and local level details\cite{chen2022scaling} or reconstruct randomly masked regions in patches\cite{he2022masked}, all of which originated from SSL techniques developed for natural images. Moreover, the similarity between patches from the same WSI or sharing the same label can be leveraged, for example, by matching semantically or spatially close patches to similar embeddings\cite{abbet2020divide,wang2022transformer}. SSL is becoming the norm for pre-training feature extractors, with a growing body of work showing that SSL strategies can provide more disentangled, robust and generalizable representations than supervised learning methods\cite{chen2022scaling, wang2022transformer, ciga2022self}. Furthermore, SSL is being explored for learning WSI embeddings, for example, by leveraging the hierarchical nature of histology slides\cite{}. This approach is particularly promising as its implementation would simplify the training procedure to minimal model fine-tuning\cite{chen2022self}, with the feature extractor and patch aggregator already having been trained with SSL, which would enable few-shot learning at the WSI level.

\textbf{Robustness and uncertainty.} Building generalizable models that are robust to domain shifts is essential for clinical deployment, to ensure that models trained in a controlled setting on curated datasets can deliver the same performance regardless of the environment and institution-specific factors. However, domain shifts are known to be difficult to model and detect by AI systems, which often generate overconfident predictions that fail to identify situations in which systems are likely to perform poorly — a phenomenon known as model miscalibration. These challenges are particularly prevalent for pathology data (Fig.~\ref{fig:outlook} (c)), in which biases can easily be introduced owing to a multitude of factors, such as differences between the staining protocols implemented in different institutions, the use of different scanning devices, each with a particular colour response, or different sub-populations from the training and testing data with distinctive phenotypical characteristics.

Although approaches such as stain normalization and augmentation\cite{macenko09,vahadane16,tellez19}, or site-preserving stratification\cite{howard2021The} are often used to mitigate these biases, clinical deployment of AI systems must be accompanied by safeguards to monitor model behaviour. A promising direction is equipping deep learning systems with mechanisms for uncertainty estimation, such as those based on model ensembling\cite{fort19,guo17} (inference with multiple models trained with different weight initializations) and test-time augmentation (various data augmentations to produce multiple copies of the test data). These approaches provide multiple predictions for each WSI or patient, which can be aggregated for a more reliable outcome prediction along with confidence intervals to express uncertainty\cite{pocevivciute2022generalisation}. These methods can be further used to detect out-of-distribution samples for which predictions are highly uncertain and cannot be trusted\cite{Dolezal2022Uncertainty,Limans2022Predictive}.

AI algorithms must also be robust to unbalanced datasets, an issue often encountered in rare disease detection. Without dedicated mechanisms, deep learning systems are known to be biased towards the majority class, with the risk of overestimating model performance, especially if no special consideration is paid to the choice of metrics. Furthermore, care should be taken to ensure that a system that works well at predicting clinical outcomes, such as genetic mutation or risk score, is not confounded by other clinical covariates, such as site, ethnicity or gender\cite{Coelho2020Causality}. Overall, ensuring the robustness of AI solutions will require substantial methodological advances, and, perhaps even more importantly, multi-institutional collaborations on a massive scale to build better datasets.

\clearpage
\subsubsection*{5.3 Translational considerations}

To date, few works in computational pathology have made a clinical impact, with only a single CPath system having received FDA approval and only a handful having CE marks in the European Union. The hurdles that creators of AI systems for CPath must overcome to gain regulatory approval are multifarious, and they include whole-slide scanners not being ubiquitous, which renders data acquisition non-trivial. This limitation has meant that almost all published CPath studies have been retrospective; moreover, dataset sizes are relatively small compared with the diversity within any given data type, and external validation cohorts are not universal. Therefore, the generalization gap, despite being present in many artificial intelligence (AI) subfields, is particularly acute in CPath.

Despite regulatory bodies not having extensive experience with AI software, they have recently begun adopting guidelines and conferring with AI experts, partly pushed by the substantially increasing influx of regulatory applications from radiology. For example, the FDA recently published an ``Artificial Intelligence and Machine Learning Software as a Medical Device Action Plan'' that outlines the steps the administration intends to take to regulate this class of software packages. Moreover, the FDA’s Office of Science and Engineering Laboratories operates different research programs, including the AI/ML Program in the FDA’s Center for Devices and Radiological Health. Finally, the FDA also participates in several collaborative communities involving AI experts from industry and academia, such as the Pathology Innovation Collaborative Community.

Although small in number, the systems that have gained regulatory approval span the range of CPath applications, from Paige Prostate, FDA-approved for the detection of foci suspicious for prostate cancer, to Mindpeak Breast Ki-67 HS, CE-marked for automated recognition and analysis of Ki-67 immunohistochemistry in breast cancer, to DoMore Diagnostics Histotype Px Colorectal, a CE-marked system that prognosticates from WSIs of colorectal cancer. Several other companies are actively developing CPath tools for clinical use, such as Owkin, DeePathology, Tempus, Stratipath, Visiopharm, PathAI, Ibex, and Indica Labs; however, actual clinical adoption and regulatory approval of CPath algorithms is, as yet, rare. Clinical use of such systems requires whole-slide scanners and associated informatics infrastructure to store slides; it also requires the delivery of the slides to the algorithm and that of the results to the pathologist, which can be prohibitively expensive. Other informatics hurdles include interoperability with a pathology practice’s laboratory information system and electronic medical record; in fact, although these communications sound simple in theory, in practice, they require substantial resources, especially if images and algorithm results need to be integrated with the hospital system’s radiology infrastructure.

Even after all regulatory and infrastructural hurdles have been overcome, the use of a CPath software solution is not guaranteed and depends on clinician buy-in and cost-effectiveness. Because AI systems are new to medicine, clinicians generally have little to no training on how they work and the best way to use them. This issue can create a level of distrust, especially considering that AI applications are often perceived as uninterpretable black boxes. Even after trust is built, health-care systems and individual clinicians will only use CPath tools provided there is a way to pay for them. For this purpose, cost reduction and outcome improvement following tool adoption needs to be demonstrated or, as in the case of some healthcare systems such as that of the USA, issuing specific billing codes that allow a clinician to be reimbursed for using that tool.

A solid demonstration of clinical benefit is only the first step on the long road to clinical adoption, but the academic interest and strong financial incentives from investors surrounding the CPath ecosystem look promising. Furthermore, several pharmaceutical companies are setting up laboratories to study how CPath can improve the drug discovery workflow, including the development of computational biomarkers that would be clinically adopted to predict response (or non-response) to treatment. The increasing involvement of all stakeholders, from regulatory bodies to clinicians, indicates that the number of CPath systems achieving regulatory approval and, ultimately, clinical adoption is only likely to increase\cite{lennerz2022unifying}.

\section{Conclusion}

The elements that have accelerated progress in the field of CPath show no signs of fading, from advances in digitizing the routine clinical workflow to progress in AI. Although the promises of automating labour-intensive manual work and reducing diagnosis variability between pathologists are already enticing, CPath has just as much potential to establish itself as a major component of research in pathology by enabling the discovery of morphological biomarkers that reflect molecular alterations, patient prognosis and the prediction of response to treatment. For CPath to have an effect on both clinical practice and biomedical research, it must work with two goals in mind: building large-scale, diverse and multimodal cohorts, and advancing representation learning of tissue with better deep learning frameworks. These goals are unlikely to be achievable within the confines of a single organization. Only through the concerted effort of multi-institutional data collection initiatives, open-source software packages and continued technical inspiration from advancements in computer vision and AI research can the said goals be reached.

\end{spacing}

\clearpage
\begin{center}
    \textbf{Pathology concepts}
\end{center}
\noindent\textbf{Whole slide image (WSI):} An image obtained by digitizing at high-resolution a glass slide using a scanner. \\
\textbf{Digital pathology:} A set of tools and systems for the acquisition, management, and diagnosis of pathology glass slides in a digital setting.\\
\textbf{Hematoxylin and eosin (H\&E) staining:} The reference stain for histological analysis of tissues for visualization of cell nuclei (in purple) with extracellular information and cytoplasm (in pink). \\
\textbf{Immunohistochemistry (IHC):} A collection of staining techniques to identify the presence of specific antigens within cells (or markers).\\
\andrew{\textbf{Tumor microenvironment (TME):} The environment surrounding a tumor (composed of normal cells, molecules, and blood vessels) that interacts with the tumor by influencing its growth and spread. }\\
\textbf{Multiplex imaging:} Imaging technique for simultaneous testing of several markers in a tissue. \\
\textbf{Computational pathology (CPath):} Computational methods based on the microscopic analysis of cells and tissues for the study of disease.
\begin{center}
    \textbf{AI concepts}
\end{center}
\andrew{\textbf{Deep Learning}:
A subdivision of machine learning where a functional relationship is learned between raw input data and some label (e.g., for image classification or segmentation). The input-output relationship is parametrized by an artificial neural network composed of a series of linear projections and non-linear activations.
}\\
\textbf{Embeddings (Representations):} A compressed and informative low-dimensional representation of a high-dimensional raw input.\\
\andrew{\textbf{Segmentation:} The pixel-level delineation of the constituents of an image. In \textit{semantic segmentation}, each pixel is associated with a category, while \textit{instance segmentation} also 
identifies individual objects within each category.}\\
\textbf{Convolutional neural network (CNN)}: A class of neural networks designed to efficiently learn from images (leveraging parameter-sharing) through the parametrization of trainable convolutional filters repetitively applied throughout the whole input. \\
\textbf{Graph neural network (GNN)} A class of neural networks designed to learn on graph-structured data. A GNN iteratively updates and aggregates information from each node's neighbor to contextualize its representation. \\
\andrew{\textbf{Transformer}: A deep learning architecture that uses self-attention to learn a differential weighting of the relevance of each element in the input data. Vision transformers extend this principle to images.}\\
\textbf{Supervised learning}: A learning paradigm in which a labeled training dataset of inputs (e.g., WSIs) and targets (e.g., cancer subtype) pairs is used to train a neural network, such that the algorithm can successfully predict a target of the test input.\\
\textbf{Self-supervised learning}: A learning paradigm where the training signal is obtained from the input itself by leveraging its underlying structure. SSL can be trained on orders of magnitude more data than SL, as no labels are required. \\
\textbf{Early and late fusion:} In applications with multiple data modalities or magnification levels, early fusion methods use a multi-modal encoder to merge the data locally; while late fusion methods create modality-level embeddings with uni-modal encoders that are then fused for prediction.\\

\begin{nolinenumbers}
\section*{References}

\bibliographystyle{nature}
\bibliography{review}
\end{nolinenumbers}

\end{document}